\documentclass[12pt]{article}

\usepackage{amssymb,amsmath}

\usepackage{graphicx}
\DeclareGraphicsExtensions{.pdf,.png,.jpg}

 \catcode`\@=11 \@addtoreset{equation}{section}\catcode`\@=12
\newcommand{\R}{ {\mathbb R} }

\begin{document}

 \begin{center}

 \large \bf Stable  exponential cosmological  solutions with zero variation of $G$  in the Einstein-Gauss-Bonnet  model 
 with a $\Lambda$-term
  \end{center}

 \vspace{0.3truecm}

 \begin{center}

  K. K. Ernazarov$^{1}$ and V. D. Ivashchuk$^{1,2}$ 

\vspace{0.3truecm}

 \it $^{1}$Institute of Gravitation and Cosmology,
 RUDN University, 6 Miklukho-Maklaya ul.,
 Moscow 117198, Russia

 \it $^{2}$ Center for Gravitation and Fundamental Metrology,
 VNIIMS, 46 Ozyornaya ul., Moscow 119361, Russia

\end{center}

\begin{abstract}

A $D$-dimensional  gravitational model with a Gauss-Bonnet term and the cosmological term $\Lambda$ is considered.
By assuming diagonal cosmological metrics, we find, for certain fine-tuned $\Lambda$, a class of solutions with  exponential time dependence of two scale factors, governed by two Hubble-like parameters $H >0$ and $h < 0$, corresponding to factor spaces of dimensions $m > 3$ and $l > 1$, respectively, with $(m,l) \neq  (6,6), (7,4), (9,3)$ and $D = 1 + m + l$. Any of these solutions  describes an exponential expansion of  3-dimensional subspace with Hubble parameter $H$ and  zero variation of the effective gravitational constant $G$. We  prove  the stability of these solutions in a class of cosmological solutions with diagonal  metrics.

\end{abstract}


\section{Introduction}

In this paper we consider $D$-dimensional gravitational model
with Gauss-Bonnet term and cosmological term $\Lambda$.  
The so-called Gauss-Bonnet term appeared in string theory as a correction to the (Fradkin-Tseytlin) effective action \cite{Zwiebach}-\cite{MTs}.

 We note that at present the Einstein-Gauss-Bonnet (EGB) gravitational model and  its modifications, 
see   \cite{Ishihara}-\cite{Ivas-16-2} and refs. therein,  are intensively studied in  cosmology,
e.g. for possible explanation  of  accelerating  expansion of the Universe which follow from supernovae (type Ia) observational data \cite{Riess,Perl,Kowalski}.

Here we  deal with the cosmological solutions with diagonal metrics  governed by $n >3$ scale factors
depending upon one variable, which is the synchronous time variable. We restrict ourselves by the
solutions with exponential dependence of scale factors and present 
a class of such solutions with  two scale factors, governed by two Hubble-like parameters $H >0$ and $h < 0$, which correspond to factor spaces of dimensions $m > 3$ and $l > 1$, respectively, with $D = 1 + m + l$ and $(m,l) \neq  (6,6), (7,4), (9,3)$.
Any of  these solutions  describes   an exponential expansion  of $3d$ subspace with Hubble parameters $H > 0$ \cite{Ade}  and has a  constant volume factor  of  $(m - 3 + l)$-dimensional internal space,
which implies zero variation of the effective gravitational constant $G$ either in Jordan or Einstein frame \cite{RZ-98,I-96}, see also  \cite{BIM,Mel,IvMel-14} and refs. therein. These solutions satisfy the most severe restrictions on  variation of $G$ \cite{Pitjeva}.

We  study   the stability of these solutions in a class of cosmological solutions with diagonal metrics by using results of  refs. \cite{ErIvKob-16,Ivas-16} (see also approach of ref. \cite{Pavl-15}) and show  that all solutions, presented here, are stable. 
It should be noted that  two  special solutions  for $D = 22, 28$ and $\Lambda = 0$ were found earlier in ref.  \cite{IvKob}. In ref. \cite{ErIvKob-16} it was proved that these solutions are stable.
Another set of six stable  exponential solutions: five in dimensions $D = 7, 8, 9, 13$ and two - for $D = 14$, were considered recently in \cite{Ivas-16-2}.

\section{The cosmological model}

The action of the model reads
\begin{equation}
  S =  \int_{M} d^{D}z \sqrt{|g|} \{ \alpha_1 (R[g] - 2 \Lambda) +
              \alpha_2 {\cal L}_2[g] \},
 \label{1}
\end{equation}
where $g = g_{MN} dz^{M} \otimes dz^{N}$ is the metric defined on
the manifold $M$, ${\dim M} = D$, $|g| = |\det (g_{MN})|$, $\Lambda$ is
the cosmological term, $R[g]$ is scalar curvature,
$${\cal L}_2[g] = R_{MNPQ} R^{MNPQ} - 4 R_{MN} R^{MN} +R^2$$
is the standard Gauss-Bonnet term and  $\alpha_1$, $\alpha_2$ are
nonzero constants.

We consider the manifold
\begin{equation}
   M = \R  \times   M_1 \times \ldots \times M_n 
   \label{2.1}
\end{equation}
with the metric
\begin{equation}
   g= - d t \otimes d t  +
      \sum_{i=1}^{n} B_i e^{2v^i t} dy^i \otimes dy^i,
  \label{2.2}
\end{equation}
where   $B_i > 0$ are arbitrary constants, $i = 1, \dots, n$, and
$M_1, \dots,  M_n$  are one-dimensional manifolds (either $\R$ or $S^1$)
and $n > 3$.

Equations of motion for the action (\ref{1}) 
give us the set of  polynomial equations \cite{ErIvKob-16}
\begin{eqnarray}
  G_{ij} v^i v^j + 2 \Lambda
  - \alpha   G_{ijkl} v^i v^j v^k v^l = 0,  \label{2.3} \\
    \left[ 2   G_{ij} v^j
    - \frac{4}{3} \alpha  G_{ijkl}  v^j v^k v^l \right] \sum_{i=1}^n v^i \nonumber
    - \frac{2}{3}   G_{ij} v^i v^j  +  \frac{8}{3} \Lambda = 0,
   \label{2.4}
\end{eqnarray}
$i = 1,\ldots, n$, where  $\alpha = \alpha_2/\alpha_1$. Here
\begin{equation}
G_{ij} = \delta_{ij} -1, \qquad   G_{ijkl}  = G_{ij} G_{ik} G_{il} G_{jk} G_{jl} G_{kl}
\label{2.4G}
\end{equation}
are, respectively, the components of two  metrics on  $\R^{n}$ \cite{Iv-09,Iv-10}. 
The first one is a 2-metric and the second one is a Finslerian 4-metric.
For $n > 3$ we get a set of forth-order polynomial  equations.

We note that for $\Lambda =0$ and $n > 3$ the set of equations (\ref{2.3}) and (\ref{2.4}) has an isotropic solution $v^1 = \ldots = v^n = H$ only if $\alpha  < 0$ \cite{Iv-09,Iv-10}.
This solution was generalized in \cite{ChPavTop} to the case $\Lambda \neq 0$.

It was shown in \cite{Iv-09,Iv-10} that there are no more than
three different  numbers among  $v^1,\dots ,v^n$ when $\Lambda =0$. This is valid also
for  $\Lambda \neq 0$ if $\sum_{i = 1}^{n} v^i \neq 0$  \cite{Ivas-16}.

\section{Solutions with constant $G$}

In this section we present a class of solutions to the set of equations (\ref{2.3}), (\ref{2.4}) of the following form
\begin{equation}
  \label{3.1}
   v =(\underbrace{H,H,H}_{``our" \ space},\underbrace{\overbrace{H, \ldots, H}^{m-3}, 
   \overbrace{h, \ldots, h}^{l}}_{internal \ space}).
\end{equation}
where $H$ is the Hubble-like parameter corresponding  
to an $m$-dimensional factor space with $m > 3$ and $h$ is the Hubble-like parameter 
corresponding to an $l$-dimensional factor space, $l>1$. We split the $m$-dimensional  
factor space into the  product of two subspaces of dimensions $3$ and $m-3$, respectively. 
The first one is identified with ``our" $3d$ space while the second one is considered as 
a subspace of $(m-3 +l)$-dimensional internal space.
 
We put 
\begin{equation}
  \label{3.2a}
   H > 0 
\end{equation}
for a description of an accelerated expansion of a
$3$-dimensional subspace (which may describe our Universe) and also put
\begin{equation}
  \label{3.2}
(m-3) H + lh  = 0
\end{equation}
for a  description of a zero variation of the effective gravitational constant $G$.

We remind (the reader) that  the effective gravitational constant $G = G_{eff}$ in the Brans-Dicke-Jordan (or simply Jordan) frame \cite{RZ-98} (see also \cite{I-96})
is proportional to the inverse volume scale factor
of the internal space, see \cite{BIM,IvMel-14} and references therein.

{\bf Remark}. {\em   Due to  ansatz (\ref{3.1}) ``our'' 3d space expands (isotropically) with 
Hubble parameter $H$ and $(m -3)$-dimensional part of internal space expands (isotropically) with the same Hubble parameter $H$ too. To avoid possible puzzles with separation of  these two subspaces, we consider for physical applications (in our epoch) the internal space to be compact one, i.e. we put in (\ref{2.1}) $M_4 = \ldots = M_n = S^1$. We also put the internal scale factors corresponding to present time $t_0$ :       $a_j (t_0) = B_j^{1/2} \exp(v^j t_0) $, $j =4, \ldots, n$, (see (\ref{2.2})) to be small enough in comparison with the scale factor of ``our'' space for $t = t_0$: $ a (t_0)  = B^{1/2} \exp(H t_0) $, where  $B_1 = B_2 = B_3 = B$.}

According to the ansatz (\ref{3.1}),  the $m$-dimensional factor space is expanding with the Hubble parameter $H >0$, while the $l$-dimensional factor space is contracting with the Hubble-like  parameter $h < 0$.

It was shown in \cite{Ivas-16} (for more general prescription 
see also \cite{ChPavTop1}) that if we consider the ansatz 
(\ref{3.1}) with two Hubble-like parameters $H$ and $h$  
with two restrictions imposed
   \begin{equation}
   m H + lh \neq 0, \qquad  H \neq h,
   \label{3.3}
   \end{equation}
 then relations (\ref{2.3}) and (\ref{2.4}) may be reduced  to the  following set of  equations
  \begin{eqnarray}
   E =  m H^2 + l h^2 - (mH + lh)^2  + 2 \Lambda 
        - \alpha [m (m-1) (m-2) (m - 3) H^4
          \nonumber \\
       + 4 m (m-1) (m-2) l H^3 h   
       + 6 m (m-1) l (l - 1) H^2 h^2
         \nonumber \\
       + 4 m l (l - 1) (l - 2) H h^3  
       + l (l - 1) (l - 2) (l - 3) h^4] = 0, \quad
         \label{3.4}   \\
  Q =  (m - 1)(m - 2)H^2 + 2 (m - 1)(l - 1) H h 
        \nonumber \\ 
       + (l - 1)(l - 2)h^2 = - \frac{1}{2 \alpha}.
     \label{3.5}
  \end{eqnarray}
The restrictions (\ref{3.3}) are satisfied for our ansatz (\ref{3.2a}) and (\ref{3.2}).         

Using equation (\ref{3.2}),  (\ref{3.5}) we get for $m > 3$ and $l > 1$ 
\begin{equation}
H   =   l  (- 2 \alpha P)^{-1/2},  \quad  h = - (m-3) H/l <  0,
 \label{3.6}
\end{equation}
where 
\begin{equation}
P  =  P(m,l)  =  (l + m - 3) ( (5 - m) l + 2m  - 6)  \neq 0 
 \label{3.7} 
\end{equation}
and 
\begin{equation}
\alpha P < 0. 
 \label{3.8} 
\end{equation}

The substitution of relation (\ref{3.6}) into (\ref{3.4})
gives us 
\begin{equation}     
\Lambda = \Lambda(m,l) =  l ( -4 \alpha P)^{-1} ( M - (1/2) P^{-1} R)
 \label{3.8} 
\end{equation}
where
\begin{equation}
M = M(m,l) = (9 - m) l - (m - 3)^2
\label{3.9}                                                       
\end{equation}
and
\begin{eqnarray}
  R = R(m,l) = - 3  m (m - 1) (m - 2) (m - 3) l^3  
             + 6 m (m -1) (m - 3)^2 l^2 (l - 1) 
              \nonumber \\                                                                               
    - 4 m (m - 3)^3 l (l-1)(l - 2) 
    + (m - 3)^4  (l - 1) (l - 2)(l - 3). \qquad \qquad
   \label{3.10}                  
\end{eqnarray}

It may be verified that equality $P(m,l) = 0$ takes place for  
$(m,l) = (9,3), (7,4), (6,6)$. 

The domains with different signs of $P =P(m,l)$ and $\alpha$ are depicted
in Figure 1, where we  enlarged our setup by adding  the case $m = 3$, which gives a solution with $h =0$. For more general solution with $m \geq 3$ and $h =0$ see also \cite{Ivas-16}.

\begin{figure}[h]
\center{\includegraphics[scale=0.35]{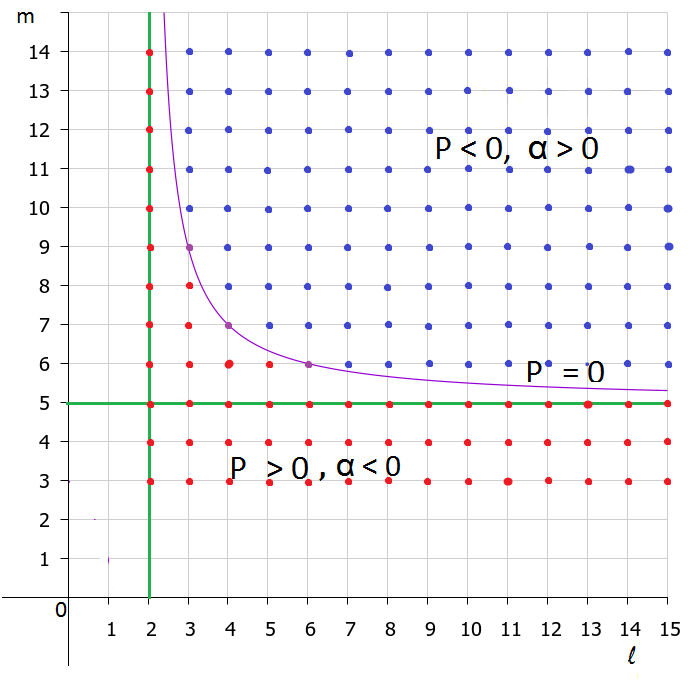}}
\caption{The domains with different signs of $P =P(m,l)$ and $\alpha$.
Red points correspond to $P > 0$ and $\alpha < 0$, while blue points
correspond to $P < 0$ and $\alpha > 0$.}
\label{fig1}
\end{figure}

Relation $P(m,l)  > 0$,  or $\alpha < 0$, takes place in the following cases 
 \begin{eqnarray}  
 (m \geq 3; l =2), \qquad (m = 3, 4, 5 ; l \geq 3), 
 \nonumber   \\ 
 (m = 6,7, 8; l = 3), \qquad ( m = 6 ; l = 4, 5). 
 \label{3.11} 
\end{eqnarray} 

Relation $P(m,l)  < 0$  or $\alpha > 0$  is valid in the following cases
 \begin{eqnarray}  
 (m \geq 10; l \geq 3), \qquad (m = 8, 9; l \geq 4),  
  \nonumber   \\ 
 (m = 7; l \geq 5), \qquad (m = 6 ; l \geq 7).
 \label{3.12} 
\end{eqnarray}

The domains with different signs of $\Lambda = \Lambda(m,l)$ are depicted
in Figure 2.

For fixed $l  > 2$ we get the asymptotic relation
 \begin{equation}
 \Lambda(m,l) \sim  \frac{l}{8 \alpha (l - 2)} 
   \label{3.13} 
  \end{equation}
as $m \to + \infty$. 

For fixed $m \geq 3$,  $m \neq 5$ and  $m \neq 9$, we obtain
 \begin{equation}
 \Lambda(m,l) \sim \frac{(m - 9) (m+1)}{8 \alpha (m - 5)^2} 
 \label{3.14}
 \end{equation}
as $l \to + \infty$, while
\begin{equation}
\Lambda(5,l)  \sim   - \frac{3 l^2}{16 \alpha}  \to + \infty, 
\end{equation}
and
\begin{equation}
\Lambda(9,l)  \sim   - \frac{9}{4 \alpha l} \to 0, 
\end{equation}
 as  $l \to + \infty$.

\begin{figure}[h]
\center{\includegraphics[scale=0.35]{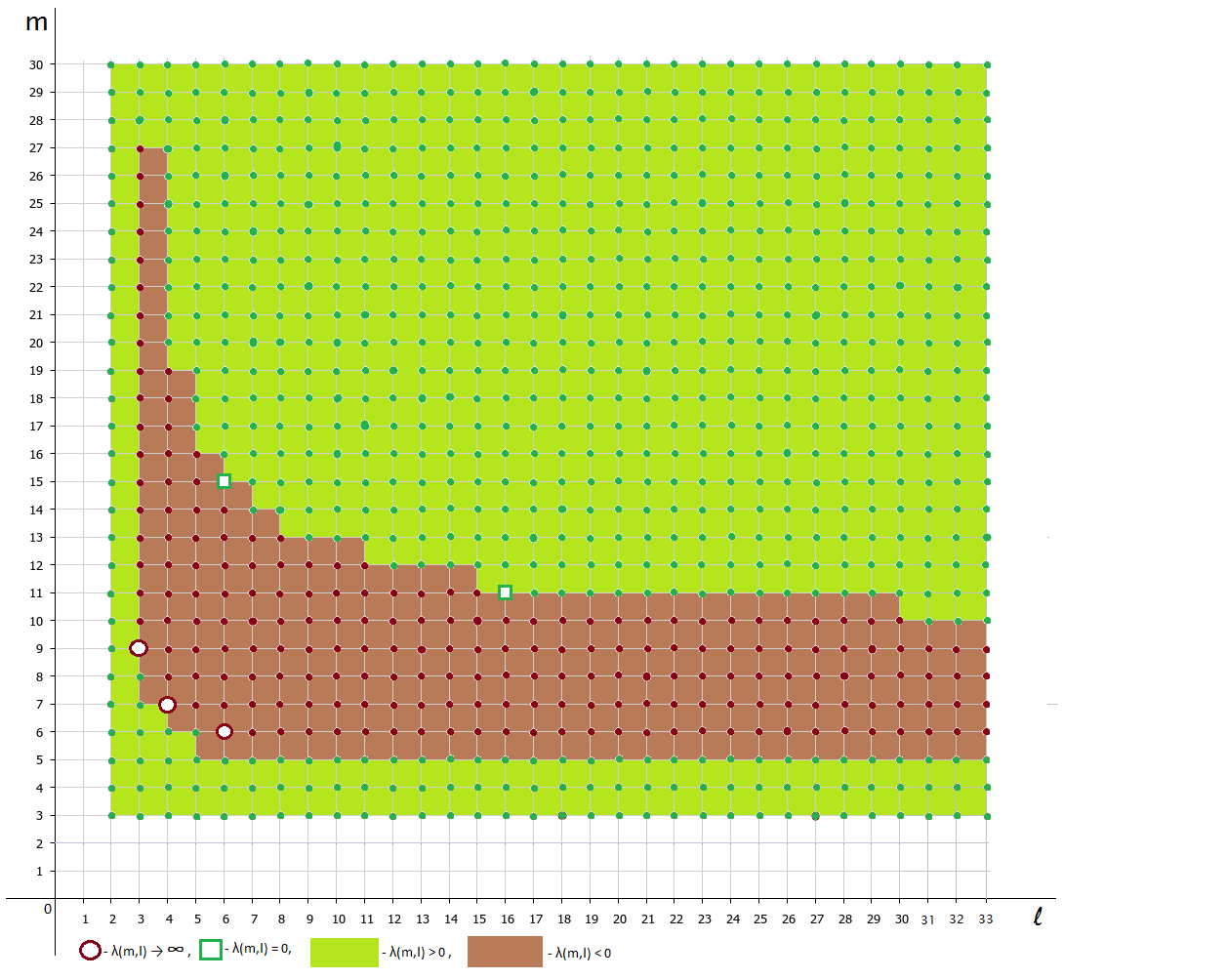}}
\caption{The domains with different signs of $\Lambda = \Lambda(m,l)$.
Green points correspond to  $\Lambda >0$ and brown points indicate $\Lambda < 0$.} 
\label{fig2}
\end{figure}

For   $m=11$,  $l=16$ and $\alpha = 1$ we get $\Lambda =0$,
$H= \frac{1}{\sqrt{15}}$ and $h= -\frac{1}{2 \sqrt{15}}$.
This solution was found in \cite{ErIvKob-16}. 
For $m=15$, $l=6$ and $\alpha = 1$ we are led to another 
solution from \cite{IvKob} with $\Lambda =0$, $H=\frac{1}{6}$ and $h=-\frac{1}{3}$. 
It was proved in \cite{ErIvKob-16} that these two solutions are stable.

\section{Stability analysis}

Here, as in \cite{ErIvKob-16,Ivas-16}, we deal with exponential solutions (\ref{2.2})
with non-static volume factor, which is proportional to $\exp(\sum_{i = 1}^{n} v^i t)$, i.e. we put
\begin{equation}
  K = K(v) = \sum_{i = 1}^{n} v^i \neq 0.
  \label{4.1}
\end{equation}

We  put the following restriction 
\begin{equation}
  ({\rm R }) \quad  \det (L_{ij}(v)) \neq 0.
  \label{4.2}
\end{equation}
on the matrix 
\begin{equation}
L =(L_{ij}(v)) = (2 G_{ij} - 4 \alpha G_{ijks} v^k v^s).
   \label{4.1b}
 \end{equation}

For general cosmological setup with the metric 
\begin{equation}
 g= - dt \otimes dt + \sum_{i=1}^{n} e^{2\beta^i(t)}  dy^i \otimes dy^i,
 \label{4.3}
\end{equation}
we have  the (mixed) set of algebraic and differential equations \cite{Iv-09,Iv-10}
\begin{eqnarray}
     E = G_{ij} h^i h^j + 2 \Lambda  - \alpha G_{ijkl} h^i h^j h^k h^l = 0,
         \label{4.3.1} \\
         Y_i =  \frac{d L_i}{dt}  +  (\sum_{j=1}^n h^j) L_i -
                 \frac{2}{3} (G_{sj} h^s h^j -  4 \Lambda) = 0,
                     \label{4.3.2a}
          \end{eqnarray}
where $h^i = \dot{\beta}^i$,           
 \begin{equation}
  L_i = L_i(h) = 2  G_{ij} h^j
       - \frac{4}{3} \alpha  G_{ijkl}  h^j h^k h^l  
       \label{4.3.3},
 \end{equation}
 $i = 1,\ldots, n$.

It was proved in \cite{Ivas-16} that a fixed point solution
$(h^i(t)) = (v^i)$ ($i = 1, \dots, n$; $n >3$) to eqs. (\ref{4.3.1}), (\ref{4.3.2a})
obeying restrictions (\ref{4.1}), (\ref{4.2}) is  stable under perturbations
\begin{equation}
 h^i(t) = v^i +  \delta h^i(t), 
\label{4.3h}
\end{equation}
 $i = 1,\ldots, n$,  (as $t \to + \infty$)  if
\begin{equation}
  K(v) = \sum_{k = 1}^{n} v^k > 0
 \label{4.1a}
\end{equation}
and  it is unstable (as $t \to + \infty$) if $K(v) = \sum_{k = 1}^{n} v^k < 0$.

We remind the reader that the  perturbations $\delta h^i$ 
obey (in linear approximation) the following set of  equations
\cite{ErIvKob-16,Ivas-16}
 \begin{eqnarray}
   C_i(v) \delta h^i = 0, \label{4.2C} \\
   L_{ij}(v) \delta \dot{h}^j =  B_{ij}(v) \delta h^j,
  \label{4.3LB}
 \end{eqnarray}
  where
 \begin{eqnarray}
 C_i(v)  =  2 v_{i} - 4 \alpha G_{ijks}  v^j v^k v^s, \label{4.3.4} \\
 L_{ij}(v) = 2 G_{ij} - 4 \alpha G_{ijks} v^k v^s,
    \label{4.3.5} \\
 B_{ij}(v) = - (\sum_{k = 1}^n v^k)  L_{ij}(v) - L_i(v) + \frac{4}{3} v_{j},
                     \label{4.3.6}
 \end{eqnarray}
 $v_i = G_{ij} v^j$,  $L_i(v) =  2 v_{i} - \frac{4}{3} \alpha  G_{ijks}  v^j v^k v^s$
 and $i,j,k,s = 1, \dots, n$.

It was proved in ref. \cite{Ivas-16} that  the set of linear equations 
on  perturbations (\ref{4.2C}), (\ref{4.3LB})  has the following solution
   \begin{eqnarray}
       \delta h^i = A^i \exp( - K(v) t ),
       \label{4.16}  \\
         \sum_{i =1}^{n} C_i(v)  A^i =0,
         \label{4.16A}
   \end{eqnarray}
    $i = 1, \dots, n$, when restrictions (\ref{4.1}), (\ref{4.2}) are imposed.    

It was shown in  \cite{Ivas-16} that  for  the vector $v$ from  (\ref{3.1}), obeying
 relations (\ref{3.3}), the matrix $L$ has a block-diagonal form
\begin{equation}
 (L_{ij}) = {\rm diag}(L_{\mu \nu}, L_{\alpha \beta} ),
 \label{4.5}
\end{equation}
where
\begin{eqnarray}
  L_{\mu \nu} =  G_{\mu \nu} (2 + 4 \alpha S_{HH}),
 \label{4.6HH}     \\
  L_{\alpha \beta} = G_{\alpha \beta} (2 + 4 \alpha S_{hh}) 
  \label{4.6hh}
\end{eqnarray}
and
\begin{eqnarray}
  S_{HH} =  (m-2)(m -3) H^2  + 2(m-2)lHh + l(l - 1)h^2 ,
  \label{4.7}   \\
  S_{hh} = m(m-1)H^2  + 2m(l - 2)Hh+ (l- 2)(l- 3)h^2.
  \label{4.8}
\end{eqnarray}

The matrix (\ref{4.5}) is invertible if and only if  $m > 1$,  $l > 1$ and
\begin{eqnarray}
 S_{HH} \neq - \frac{1}{2 \alpha}, \label{4.9a}
  \\
 S_{hh} \neq - \frac{1}{2 \alpha},
 \label{4.9b}
\end{eqnarray}
since the matrices  $(G_{\mu \nu}) = (\delta_{\mu \nu} -1 )$ and
$(G_{\alpha \beta}) = (\delta_{\alpha \beta} - 1)$ are invertible only if  $m > 1$ and $l > 1$.

The first condition (\ref{4.1a}) is obeyed for the solutions under consideration
since due to (\ref{3.2}) we get  $K(v) = 3H > 0$ \cite{Ivas-16}.

Now, let us prove that inequalities (\ref{4.9a}), (\ref{4.9b}) are satisfied. 

Let us suppose that  (\ref{4.9a}) is not satisfied, i.e.  
$S_{HH} = - \frac{1}{2 \alpha}$.  
Then using (\ref{3.5}) we get  
\begin{equation}
S_{HH} - Q = - 2 (H - h) ((m-2) H + (l -1) h ) = 0, 
\label{4.10a}                                            
\end{equation}
which implies  due to $H - h \neq 0$ (see  (\ref{3.6})),         
\begin{equation}
   (m-2) H + (l -1) h = 0.
\label{4.11a}                                                        
\end{equation}
The substitution of $h = - (m - 3)H/l$ into (\ref{4.11a}) gives us       
 $(l  + m - 3) H   = 0$, which is in contradiction with inequalities
 $m > 3$, $l > 1$ and $H > 0$. This contradiction proves the inequality
(\ref{4.9a}).     
      
 Now we  suppose that  (\ref{4.9b}) is not valid,
 i.e. $S_{hh} = - \frac{1}{2 \alpha}$. Then using (\ref{3.5}) we get  
 \begin{equation}
 S_{hh} - Q =  - 2 (h - H)  ((l - 2) h + (m -1) H ) = 0, 
 \label{4.10b}                                            
 \end{equation}
 which implies  due to $H - h \neq 0$         
 \begin{equation}
    (l - 2) h + (m -1) H = 0.
 \label{4.11b}                                                        
 \end{equation}
 Substituting  $h = - (m - 3)H/l$ into (\ref{4.11b}) implies the  relation
  $2 ( l + m - 3) H  = 0$, which contradicts the relations
  $m > 3$, $l > 1$ and $H > 0$. The contradiction proves the inequality
 (\ref{4.9b}).         

Thus, we proved that relations (\ref{4.9a}) and (\ref{4.9b}) are valid, hence the
restrictions (\ref{4.2}) are satisfied for our solutions. 
This completes the proof of stability of the solutions under consideration.

\section{Conclusions}

We have considered the  $D$-dimensional  Einstein-Gauss-Bonnet (EGB) model
with the $\Lambda$-term and two constants $\alpha_1$ and $\alpha_2$.  By using the  ansatz with diagonal  cosmological  metrics, we have found, for certain  $\Lambda = \Lambda(m,l)$ and  $\alpha = \alpha_2 / \alpha_1 $, a class of solutions with  exponential time dependence of two scale factors, governed by two Hubble-like parameters $H >0$ and $h < 0$, corresponding to submanifolds of dimensions $m > 3$ and $l > 1$, respectively, with $(m,l) \neq  (6,6), (7,4), (9,3)$ and $D = 1 + m + l$.
Here $m > 3$ is the dimension  of the expanding subspace and $l > 1$
is the dimension of  contracting  one.

Any of these solutions describes an exponential expansion of ``our'' 3-dimensional subspace with
the Hubble parameter $H > 0$ and anisotropic behaviour of $(m-3+ l)$-dimensional internal space:
expanding in $(m-3)$ dimensions (with Hubble-like parameter $H$) and contracting in $l$ dimensions
(with Hubble-like parameter $h < 0$).
Each solution   has a constant volume factor of internal space and hence it describes
zero variation of the  effective gravitational constant $G$.  By using results of ref. \cite{Ivas-16}
we  have proved that all these solutions  are stable as  $t \to + \infty$.


 {\bf Acknowledgments}

This paper was funded by the Ministry of Education and Science of the Russian Federation
in the Program to increase the competitiveness of Peoples' Friendship University
(RUDN University) among the world's leading research and education centers in the 2016-2020
and  by the  Russian Foundation for Basic Research,  grant  Nr. 16-02-00602.

\end{document}